\documentclass[apj]{emulateapj}

\begin{document}


\title{Selection bias in the $M_\bullet-\sigma$ and $M_\bullet-L$ 
       correlations and its consequences}

\author{Mariangela Bernardi\altaffilmark{1}, 
        Ravi K. Sheth\altaffilmark{1}, 
        Elena Tundo\altaffilmark{1,2} and
        Joseph B. Hyde\altaffilmark{1}}

\affil{}

\altaffiltext{1}{Dept. of Physics and Astronomy, University of Pennsylvania, 
                 209 South 33rd St, Philadelphia, PA 19104, U.S.A.}

\altaffiltext{2}{Dipartimento di Astronomia, Universita' di Padova, 
                 vicolo dell'Osservatorio 3/2 I-35122, Padova, Italy}

\begin{abstract}
 It is common to estimate black hole abundances by using a 
 measured correlation between black hole mass and another more 
 easily measured observable such as the velocity dispersion or 
 luminosity of the surrounding bulge.  The correlation is used to 
 transform the distribution of the observable into an estimate of 
 the distribution of black hole masses.  
 However, different observables provide different estimates:  
 the $M_\bullet-\sigma$ relation predicts fewer massive black holes 
 than does the $M_\bullet-L$ relation.  This is because the 
 $\sigma-L$ relation in black hole samples currently available 
 is inconsistent with that in the SDSS sample, from which the 
 distributions of $L$ or $\sigma$ are based:  the black hole 
 samples have smaller $L$ for a given $\sigma$ or have larger $\sigma$
 for a given $L$.  This is true whether $L$ is estimated in the 
 optical or in the NIR.  
 The $\sigma-L$ relation in the black hole samples is similarly 
 discrepant with that in the ENEAR sample of nearby early-type 
 galaxies.  This suggests that current black hole samples are 
 biased towards objects with abnormally large velocity dispersions 
 for their luminosities.  If this is a selection rather  than 
 physical effect, then the $M_\bullet-\sigma$ and $M_\bullet-L$ 
 relations currently in the literature are also biased from their 
 true values.  We provide a framework for describing the effect of 
 this bias.  We then combine it with a model of the bias to make an 
 estimate of the true intrinsic relations.  While we do not claim to have 
 understood the source of the bias, our simple model is able to 
 reproduce the observed trends.  If we have correctly modeled the 
 selection effect, then our analysis suggests that the bias in the 
 $\langle M_\bullet|\sigma\rangle$ relation is likely to be small, 
 whereas the $\langle M_\bullet|L\rangle$ relation is biased towards 
 predicting more massive black holes for a given luminosity.
 In addition, it is likely that the $M_\bullet-L$ relation is entirely 
 a consequence of more fundamental relations between $M_\bullet$ and 
 $\sigma$, and between $\sigma$ and $L$.  The intrinsic relation we 
 find suggests that at fixed luminosity, older galaxies tend to host 
 more massive black holes.  
\end{abstract}

\keywords{galaxies: elliptical --- galaxies: fundamental parameters --- 
black hole physics}

\section{Introduction}
Several groups have noted that galaxy formation and supermassive 
black hole growth should be related, and many have modeled the 
joint evolution of quasars and galaxies (e.g., Monaco et al. 2000; 
Kauffmann \& Haehnelt 2001; Granato et al. 2001; 
Cavaliere \& Vittorini 2002; Cattaneo \& Bernardi 2003; Haiman et al. 2004;
Hopkins et al. 2004; Lapi et al. 2006; Haiman et al. 2006 and references 
therein).  However, the number of black hole detections to 
date is less than fifty; this small number prevents a direct estimate 
of the black hole mass function.  On the other hand, correlations 
between $M_\bullet$ and other observables can be measured quite reliably, 
if one assumes that the $M_\bullet-$observable relation is a single 
power-law.  $M_\bullet$ is observed to correlate strongly and tightly 
with the velocity dispersion of the surrounding bulge 
(Ferrarese \& Merritt 2000; Gebhardt et al. 2000; Tremaine et al. 2002), 
as well as with bulge luminosity (McLure \& Dunlop 2002) 
and bulge stellar mass (Marconi \& Hunt 2003; H\"aring \& Rix 2004): 
in all cases, a single power law appears to be an adequate description.  

Since it is considerably easier to measure bulge properties than 
$M_\bullet$, black hole abundances are currently estimated by 
transforming the observed abundances of such secondary indicators 
using the observed correlations with $M_\bullet$ (Yu \& Tremaine 2002; 
Aller \& Richstone 2002; Marconi et al. 2004; Shankar et al. 2004; 
McLure \& Dunlop 2004).  
Unfortunately, combining the $M_\bullet-\sigma$ relation with the 
observed distribution of velocity dispersions results in an estimate 
of the number density of black holes with $M_\bullet>10^9M_\odot$ 
which is considerably smaller than the estimate based on combining 
the $M_\bullet-L$ relation with the observed luminosity function 
(Tundo et al. 2006).  

The fundamental cause of the discrepancy is that the luminosity 
and velocity dispersion functions are obtained from the SDSS 
(Blanton et al. 2003; Sheth et al. 2003), but the $\sigma-L$ 
relation in the SDSS (Bernardi et al. 2003b) is different from that 
in the black hole samples (e.g. Yu \& Tremaine 2002).  
In Section~\ref{why}, we argue that the SDSS $\sigma-L$ relation is 
{\em not} to blame, suggesting that it is the black hole samples 
which are biased.  

\begin{figure*}
\centering
 \includegraphics[width=1.9\columnwidth]{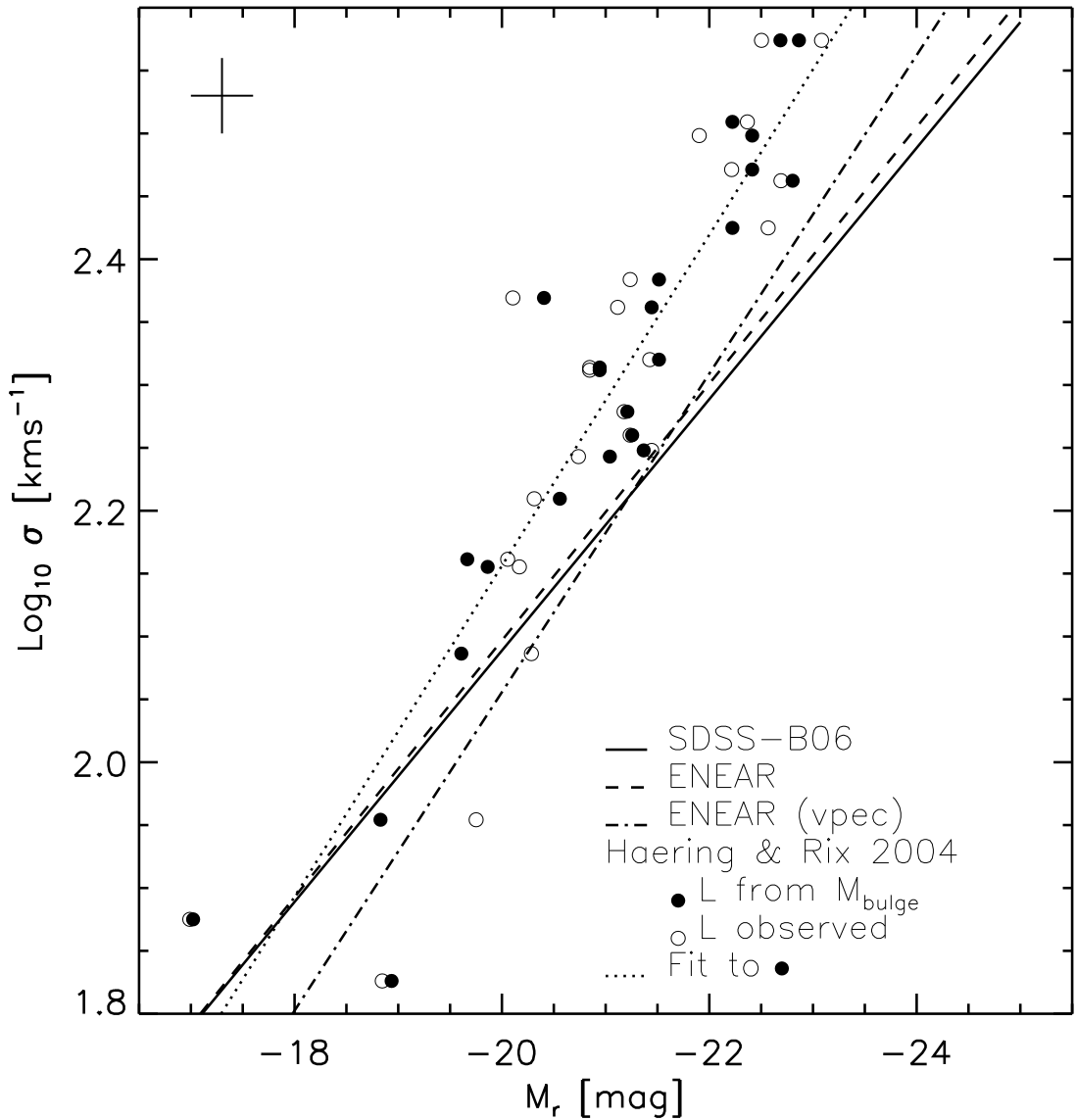}
 \caption{Comparison of the $\sigma-L$ relation in the black hole 
    sample of H\"aring \& Rix (2004) (symbols) with that from 
    local early-type galaxy samples (lines).  
    We provide two estimates of the bulge 
    luminosity for the H\"aring \& Rix compilation:  
    the open circles come from rescaling the bulge luminosity 
    to the SDSS r-band, and the filled circles from assuming a 
    mass-to-light ratio to scale the bulge masses $M_{\rm bulge}$ 
    to luminosities. Dotted line shows a fit to the filled circles.  
    Solid and short dashed lines show $\langle\sigma|L\rangle$ 
    in the SDSS-B06 and ENEAR samples.  
    Dot-dashed line shows the biased relation obtained from the 
    ENEAR sample if one fails to account for the fact that the 
    velocity dispersions played an important role in determining 
    the distances from which the luminosities were estimated.  
    The black hole compilation is clearly offset from the 
    SDSS-B06, the ENEAR and even the biased ENEAR relations.
    Appendix~A gives the fits shown, and describes exactly how the 
    black hole sample was compiled.  }
 \label{sigmaLv}
\end{figure*}

If black hole samples are indeed biased, then this bias may be 
physical or it may simply be a selection effect.  If physical, 
then only a special set of galaxies host black holes, and the 
entire approach above (of transforming the luminosity or velocity 
dispersion function) is compromised.  On the other hand, if it is a 
selection effect, then one is led to ask if the $M_\bullet-$observable 
relations currently in the literature are biased by this selection 
or not.  The main purpose of this work is to provide a framework for 
describing these biases.  
Section~\ref{biases} describes Monte-Carlo simulations 
of the selection effect.  It shows an example of the bias which 
results, and provides an analytic model of the effect.  
Our findings are summarized in Section~\ref{sowhat}, where 
estimates of the intrinsic black hole mass function $\phi(M_\bullet)$, 
and the intrinsic $M_\bullet-\sigma$ and $M_\bullet-L$ relations are 
provided.  

Throughout, we assume a spatially flat background cosmology with 
$\Omega_0=0.3$ and a cosmological constant.  All luminosities and 
masses have been scaled to $H_0 = 70$~km~s$^{-1}$~Mpc$^{-1}$.

\section{Motivation:  The $\sigma-L$ relation}\label{why}
This section shows that the $\sigma-L$ relation of early-type 
galaxies is significantly different from that in current black 
hole samples.  This difference exists whether $L$ is computed 
in a visual band, or in the near-infrared.  

\subsection{Remarks on SDSS photometry and spectroscopy}
Since 2002, when the Bernardi et al. (2003a) sample was compiled, 
a number of systematic problems with SDSS photometry and spectroscopy 
have been discovered.  The initial (SDSS-DR1 and earlier) photometry 
overestimated the luminosities and sizes of extended objects.  
Although later releases correct for these problems, they still 
overestimate the sky level in crowded fields 
(e.g. Mandelbaum et al. 2005; Bernardi et al. 2006; Lauer et al. 2006), 
and so they tend to underestimate the luminosities of extended objects 
in such fields.  This problem is particularly severe for early-type 
galaxies in nearby clusters.  In addition, the SDSS slightly 
overestimates the velocity dispersions at small $\sigma$
(the velocity dispersions in Bernardi et al. 2003a differ from
the values reported in the SDSS database and are not biased 
at small $\sigma$).  
Bernardi (2006) discusses an analysis of the $\sigma-L$ which results 
once all these systematic effects have been corrected-for:  the net 
effect does not significantly change the $\sigma-L$ relation reported 
by Bernardi et al. (2003b).  Nevertheless, the results which follow are 
based on luminosities and velocity dispersions which have been 
corrected for all these effects.  Because they are not easily 
obtained from the SDSS database, in what follows, we refer to the 
corrected SDSS sample as the SDSS-B06 sample.  

Bernardi (2006) shows that the $\sigma-L$ relation in the SDSS-B06 
sample is consistent with that in ENEAR, the definitive sample of 
nearby (within 7000~km~s$^{-1}$) early-type galaxies 
(da Costa et al. 2000; Bernardi et al. 2002; Alonso et al. 2003; 
Wegner et al. 2003), provided that one accounts for the bias 
which arises from the fact that $\sigma$ played an important 
role in determining $L$.  
Hence, the local (ENEAR) and not so local determinations 
(SDSS-B06) of the $\sigma-L$ relation are consistent with one 
another.

\subsection{The $\sigma-L$ relations in the optical}

Figure~\ref{sigmaLv} compares the correlation between $\sigma$ and 
$L$ in ENEAR and SDSS-B06 with that in the recent compilations of 
black holes by H\"aring \& Rix (2004).  The dotted line shows a 
fit to $\langle\sigma|L\rangle$ defined by the filled circles.  
Appendix~A describes this and other black hole compilations we 
have consider in the following figures.  It also gives the fits 
shown in this and the other figures.  

The solid line shows $\langle\sigma|L\rangle$ in the SDSS-B06 sample.  
The dot-dashed line shows a determination of the ENEAR $\sigma-L$ 
relation which attempts to correct for peculiar velocity effects 
in the distance estimate. This correction is known to bias the 
$\langle\sigma|L\rangle$ relation towards a steeper slope 
(Bernardi 2006).  
The short-dashed line shows the result of using the redshift as 
distance indicator when computing the ENEAR luminosities; while 
not ideal, comparison of this relation with the dot-dashed line 
provides some indication of the magnitude of the bias.  
Note in particular that using the redshift as distance indicator 
brings the ENEAR relation substantially closer to the SDSS-B06 
relation, for which the redshift is an excellent distance indicator.  

The agreement between ENEAR and SDSS-B06 suggests that the 
difference between the $\sigma-L$ relation in the black hole 
sample (dotted) and in SDSS-B06 (solid) cannot be attributed to 
problems with the SDSS determination.  

The agreement between ENEAR and SDSS-B06, and the disagreement 
with the black hole samples strongly suggests that the $\sigma-L$ 
relation in black hole samples is biased to larger $\sigma$ for 
given $L$, or to smaller $L$ for a given $\sigma$.  
Similar conclusions are obtained if we use Kormendy \& Gebhardt (2001)
or Ferrarese \& Ford (2005) compilations.
At a given luminosity, the black hole samples have $\log\sigma$ 
larger than the ENEAR or SDSS-B06 relations by $\sim 0.1$~dex. 
As a result of this difference, smaller luminosities can apparently 
produce black holes of larger mass than do their associated 
velocity dispersions (e.g., Yu \& Tremaine 2002).  
Tundo et al. (2006) show this in some detail.  

\subsection{The $\sigma-L$ relations in the near-infrared}
\begin{figure*}
\centering
\includegraphics[width=1.9\columnwidth]{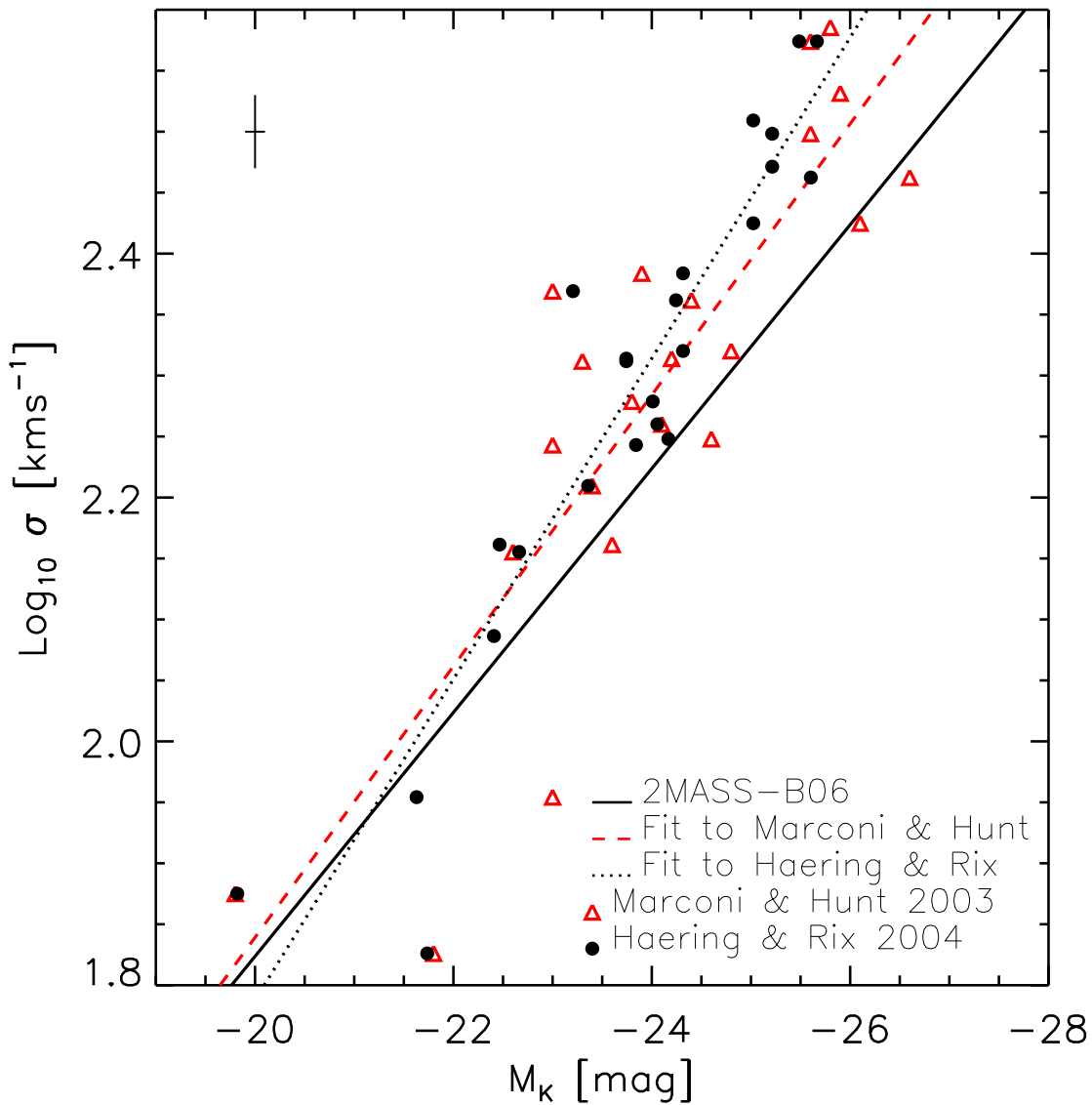}
 \caption{Same as Figure~1 but now using the $K$-band
     luminosity.  Triangles show the subset of the H\"aring \& Rix 
     sample for which $K$-band luminosity estimates are available 
     (from Marconi \& Hunt 2003).  Solid circles show the result 
     of rescaling the $M_{\rm bulge}$-based luminosities of these 
     objects to the $K$-band using $r-K = 2.8$.  Dotted and dashed 
     lines show fits to the $\langle\sigma|L_{\rm K}\rangle$ 
     relations defined by these points.  Solid line shows 
     $\langle\sigma|L_{\rm K}\rangle$ obtained by matching the 
     SDSS-B06 early-type sample to the 2MASS database, and then 
     using the $K$-band luminosities with the SDSS-B06 velocity 
     dispersions.  This fit is very similar to the solid line 
     shown in the previous panels if $r-K = 2.8$.  
     These fits are given in Appendix~A.}
 \label{sigmaLK}
\end{figure*}

We have considered the possibility that the discrepancy between 
the $\sigma-L$ relations may depend on wavelength.  
Figure~\ref{sigmaLK} presents a similar analysis, but now the 
luminosities are from the $K$-band.  In the case of the early-type 
sample, the SDSS-B06 sample was matched to 2MASS, and then the 
2MASS isophotal magnitudes were brightened by 0.2~mags to make 
them more like total magnitudes (Kochanek et al. 2001; 
Marconi \& Hunt 2003).  The solid line shows the resulting 
$\langle\sigma|L_{\rm K}\rangle$ relation.  It is very similar to 
that in the $r-$band if $r-K = 2.8$.  

Marconi \& Hunt (2003) matched their black hole compilation to 
2MASS, and then recomputed the photometry to determine $L_{\rm K}$.  
The open triangles show the $\sigma-L_{\rm K}$ relation for 
the subset of objects which are in the H\"aring \& Rix sample.  
The filled circles are obtained by taking these objects (i.e. 
those in both Marconi \& Hunt and H\"aring \& Rix), 
and then shifting the $M_{\rm bulge}$ based $r-$band luminosities 
to $K$ assuming $r-K = 2.8$.  Dashed and dotted lines show fits to 
these relations.  Comparison with the solid line (2MASS-B06) shows 
a significant offset---the black hole samples are different 
from the population of normal early-type galaxies.

\section{Biases in the $M_\bullet-\sigma$ and $M_\bullet-L$ relations}
\label{biases}
If black hole samples are indeed biased, then this bias may be 
physical or simply a selection effect.  
One is then led to ask if the relations currently in the 
literature are biased by this selection or not.  
To address this, we must be able to quantify how much of the 
$M_\bullet-\sigma$ correlation arises from the $M_\bullet-L$ 
and $L-\sigma$ relations, and similarly for the $M_\bullet-L$ 
relation.  

\subsection{Demonstration using mock catalogs}
Our approach is to make a model for the true joint distribution of 
$M_\bullet$, $L$ and $\sigma$, and to study the effects of selection 
by measuring the various pairwise correlations before and after 
applying the selection procedure to this joint distribution.  
In what follows, we will use $V = \log\, (\sigma/{\rm km~s}^{-1})$, 
and with some abuse of notation, $M_\bullet$ to denote the logarithm 
of the black hole mass when expressed in units of $M_\odot$ and $L$ 
to denote the absolute magnitude (since, in the current context, 
$M$ for absolute magnitude is too easily confused with mass).  
We assume that correlations between the logarithms of these quantities 
define simple linear relations, so our model has nine free parameters:  
these may be thought of as the slope, zero-point and scatter of 
each of the three pairwise correlations, or as the means and variances 
of $L$, $V$ and $M_\bullet$ and the three correlation coefficients
$r_{VL}$, $r_{\bullet V}$ and $r_{\bullet L}$.  

To illustrate, let $\langle V\rangle$ and $\sigma_V$ denote the mean 
and rms values of $V$, and use similar notation for the other two 
variables.  Then the mean $V$ as a function of $L$ satisfies 
\begin{equation}
 {\langle V|L\rangle - \langle V\rangle \over \sigma_V}
  = {L - \langle L\rangle\over\sigma_L}\ r_{VL},
\end{equation}
and the rms scatter around this relation is 
\begin{equation}
 \sigma_{V|L} = \sigma_V\,\sqrt{1 - r_{VL}^2}.
\end{equation}
The combination $a_{V|L}\equiv r_{VL}\,\sigma_V/\sigma_L$ is 
sometimes called the slope of the $\langle V|L\rangle$ relation, 
and $\langle V\rangle - a_{V|L}\,\langle L\rangle$ is the zero-point.  
Similarly, the slopes of $\langle M_\bullet|V\rangle$ and 
$\langle M_\bullet|L\rangle$ are 
$a_{\bullet|V}\equiv r_{\bullet V}\,\sigma_\bullet/\sigma_V$
and 
$a_{\bullet|L}\equiv r_{\bullet L}\,\sigma_\bullet/\sigma_L$
respectively.  
Notice that the zero points are trivial if one works with quantities 
from which the mean value has been subtracted.  We will do so 
unless otherwise specified.

\begin{figure*}
\centering
\includegraphics[width=1.9\columnwidth]{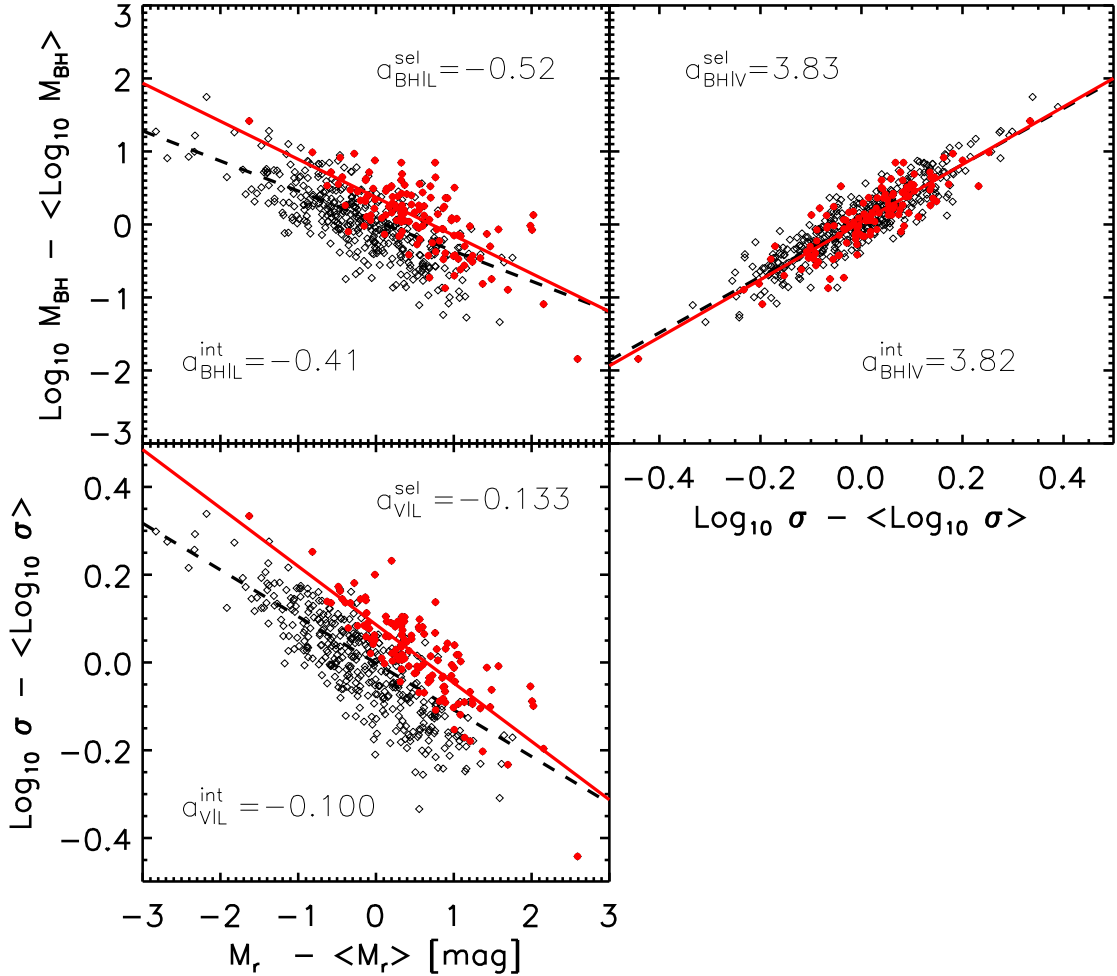}
 \caption{Comparison of the joint distribution of $M_\bullet$, $\sigma$ 
          and $L$, before (open symbols) and after (filled symbols) 
          applying a selection cut.  Dashed and solid lines show 
          the intrinsic and observed correlations.  The probability 
          of selection is assumed to increase as one moves to the 
          right of the dashed line in the bottom left panel.  }
 \label{mockRels}
\end{figure*}

Our analysis is simplified by the fact that five of the nine free 
parameters are known:  the mean and rms values of $L$ and $V$ as 
well as the correlation between $V$ and $L$ have been derived from
the SDSS-B06 sample (these values are similar to those reported
by Bernardi et al. 2003b, who also show that the distribution of $V$ 
given $L$ is Gaussian).  Hence, we use 
\begin{displaymath}
 \sigma_L = 0.84, \quad \sigma_V = 0.11,\quad {\rm and}\quad 
   r_{VL} = -0.78, 
\end{displaymath}
to make a mock catalog of the joint $V-L$ distribution in the 
SDSS-B06 early-type galaxy sample.  
Sheth et al. (2003) show explicitly that the resulting mock catalog 
provides luminosity and velocity dispersion functions which are 
consistent with those seen in the SDSS.  

We turn this into a catalog of black hole masses by assuming that, 
like $V$ at fixed $L$, the distribution 
of $M_\bullet$ at fixed $V$ and $L$ is Gaussian.  Hence, for each 
$V$ and $L$, we generate a value of $M_\bullet$ that is drawn from 
a Gaussian distribution with mean 
\begin{eqnarray}
 {\langle M_\bullet|L,V\rangle\over\sigma_\bullet} 
 &=&  {\langle M_\bullet\rangle\over\sigma_\bullet} 
      + {L-\langle L\rangle\over\sigma_L}\,
        {r_{\bullet L} - r_{\bullet V}r_{VL}\over 1 - r_{VL}^2}\nonumber\\
  &&  \quad + \quad {V-\langle V\rangle\over\sigma_V}\,
        {r_{\bullet V} - r_{\bullet L}r_{VL}\over 1 - r_{VL}^2},
\end{eqnarray}
and rms
\begin{equation}
 \sigma_{\bullet|LV} = \sigma_\bullet\, \sqrt{1 - r_{VL}^2 - r_{\bullet V}^2 
                   - r_{\bullet L}^2 + 2r_{VL}r_{\bullet L}r_{\bullet V}\over
                                    1 - r_{VL}^2}
\end{equation}
(see, e.g., Bernardi et al. 2003b).
The combination $r_{\bullet V}r_{VL}$ measures how much of the 
correlation between $M_\bullet$ and $L$ is simply a consequence 
of the correlations between $M_\bullet$ and $V$ and between 
$V$ and $L$.  The combination $r_{\bullet L}r_{VL}$ can be 
interpreted similarly.  
Notice again that the mean values of the three quantities mainly 
serve to set the zero-points of the correlations.  
Since we are working with quantities from which the mean has been 
subtracted, our model for the true relations is fixed by specifying 
the values of $\sigma_\bullet$, $r_{\bullet V}$ and $r_{\bullet L}$.  

\begin{figure*}
\centering
\includegraphics[width=1.9\columnwidth]{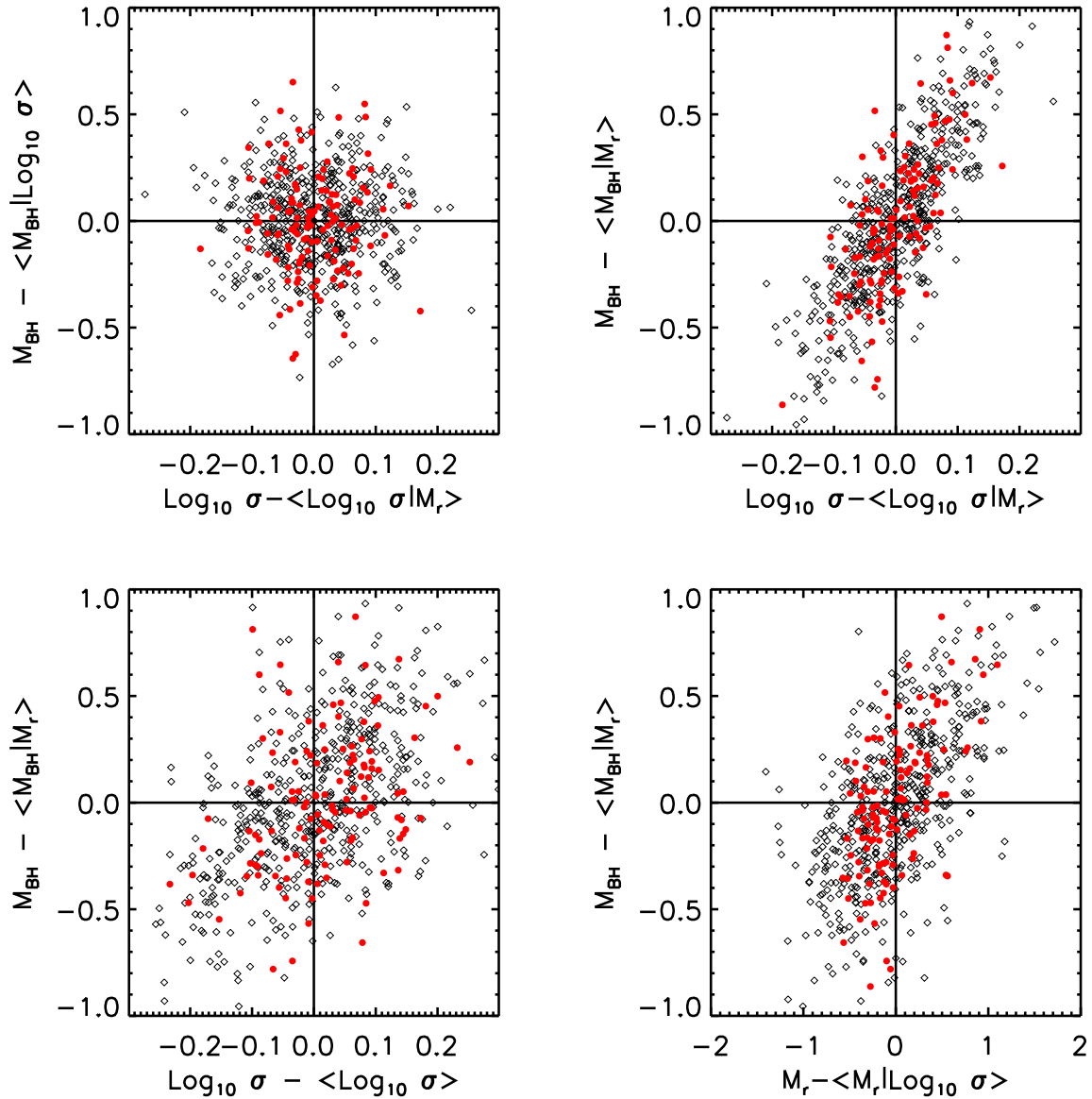}
 \caption{Residuals from the $M_\bullet-\sigma$ relation do not 
          correlate with residuals from the $\sigma-L$ correlation 
          (top left); whereas residuals from the $M_\bullet-L$ 
          relation do (top right).  They also correlate with residuals 
          from the $L-\sigma$ relation (bottom right), and weakly with 
          $\sigma$ (bottom left).  In all panels, open circles 
          show the intrinsic correlations and filled circles show 
          correlations after applying the selection cut.  
          Comparison with Figure~\ref{BHresids} shows that the 
          filled circles trace similar locii to those seen in the 
          H\"aring \& Rix (2004) compilation.}
 \label{SIMresids}
\end{figure*}

\begin{figure*}
\centering
\includegraphics[width=1.9\columnwidth]{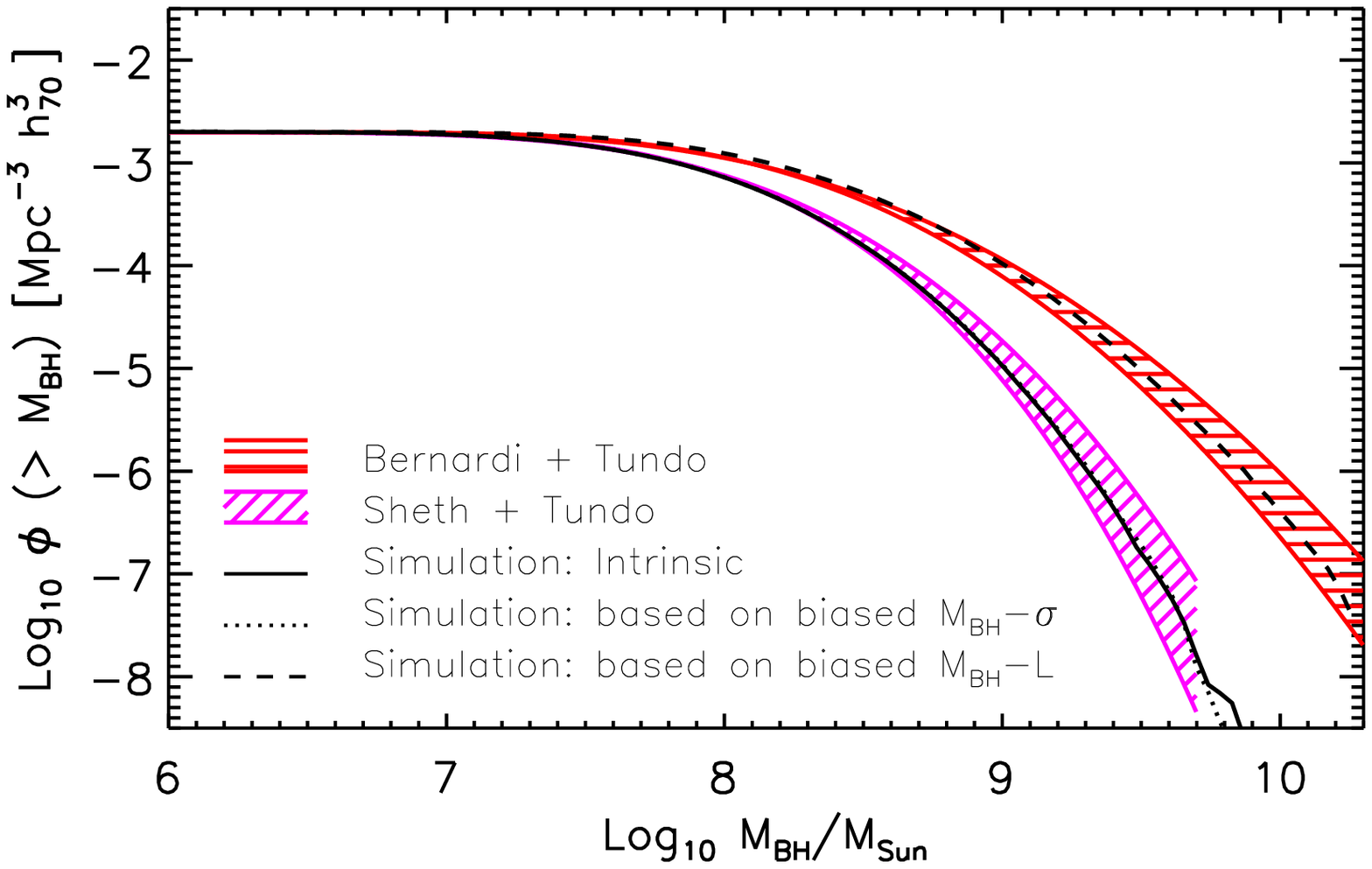}
 \caption{Comparison of the intrinsic (solid) and selection-biased 
          counts based on using velocity dispersion 
          (dotted) and luminosity (dashed) as the indicator of 
          $M_\bullet$.  Hashed regions show the abundances 
          infered by Tundo et al. (2006) by combining the 
          $M_\bullet-L$ and $M_\bullet-\sigma$ relations in 
          the compilation of H\"aring \& Rix (2004) with the 
          SDSS-based luminosity and velocity functions 
          (Bernardi et al. 2003b; Sheth et al. 2003).}
 \label{mockCounts}
\end{figure*}

To study how selection effects might have biased the observed 
$M_\bullet-\sigma$ and $M_\bullet-L$ relations we generate a mock 
galaxy+black hole catalog as described above, model the selection 
effect, and then measure the three pairwise correlations in the 
selection-biased catalog.  Figure~\ref{mockRels} shows an example 
of the result when 
\begin{displaymath}
 \sigma_\bullet = 0.49,\quad  r_{\bullet V} = 0.88,\quad 
 {\rm and}\quad r_{\bullet L} = -0.70.  
\end{displaymath}

We have tried a variety of ways to specify a selection effect.  
In Figure~\ref{mockRels}, we supposed that no objects with 
luminosities brighter than $\langle L|V\rangle$ were selected, and 
that the fraction of fainter objects which are selected is 
${\rm erf}(x/\sqrt{2})$, where $x = (L - \langle L|V\rangle)/\sigma_{L|V}$.  
The open diamonds in each panel show the objects in the original catalog, 
and the filled circles show the objects which survived the selection 
procedure.  The panels also show the slopes of the relations before 
(dashed) and after (solid) selection.  
Note that because the selection was made at fixed $V$, 
the slope of $\langle M_\bullet|V\rangle$ in the selected sample 
is not changed from its original value.  
However, $\langle M_\bullet|L\rangle$ and $\langle V|L\rangle$
are both steeper than before.  The exact values of the change 
in slope depend upon our choices for the intrinsic model, 
for reasons we describe in the next subsection.  

The parameter choices given above, with the selection described 
above, result in zero-points, slopes (given at top of each panel) 
and scatter (0.22~dex, 0.33~dex and 0.05~dex in the top left, right, 
and bottom panels, respectively) which are in excellent agreement 
with those in the H\"aring \& Rix (2004) compilation (see Appendix~A).  
Figure~\ref{SIMresids} shows that residuals from these relations 
exhibit similar correlations to those seen in the real data (compare 
Figure~\ref{BHresids}; residuals from the $M_\bullet-\sigma$ relation 
do not correlate with $\sigma$, $L$ or $L-\sigma$ residuals, so we 
have not shown these correlations here).  
This agreement is nontrivial: for example, it is more difficult to 
match all three observed correlations if the selection is based on 
$\langle V|L\rangle$ rather than $\langle L|V\rangle$.  
Changes in the parameter values of more than five percent produce 
noticable differences.  

The correlations shown in the two right hand panels of 
Figure~\ref{SIMresids} are similar in the full and in the 
bias selected sample, suggesting that the selection has not 
altered the shape of this correlation.  This suggests that the 
correlations seen in the right hand panel of Figure~\ref{BHresids} 
are real--they are not due to selection effects.  We comment in 
the final section on what these correlations might imply.  

The result of using the biased $\langle M_\bullet|L\rangle$ relation 
(both slope and scatter) in the mock luminosity function to predict 
black hole abundances is shown as the dashed line in 
Figure~\ref{mockCounts}.  It overpredicts the true abundances in 
the mock catalog, which are shown by the solid line.  The dotted 
line shows that the analogous $\sigma-$based predictor is, essentially 
unbiased.  This is not surprising in view of the results shown in  
Figure~\ref{mockRels}---the observed $\langle M_\bullet|L\rangle$ 
relation is biased towards predicting more massive black holes for 
a given luminosity, whereas the $\langle M_\bullet|\sigma\rangle$ 
is almost completely unbiased.  The hashed regions show the $L-$ 
and $\sigma-$based predictions derived by using the relations in 
the Appendix (taken from Tundo et al. 2006) with the 
Bernardi et al. (2003b) luminosity function and the Sheth et al. (2003) 
velocity function.  
They bracket the results from our mock catalog quite accurately.  
Note, however, that in our mock catalog, the true abundances are 
given by the solid line:  the much larger abundances predicted from 
the luminosity-based approach are due to the selection effect.  For 
reference, this intrinsic distribution is well-described by 
equation~(\ref{phiMbh}).  

\subsection{Toy model of the bias}
If current black hole samples only select objects which have 
$\log\sigma$ which is exactly $n$-standard deviations above the 
$\langle V|L\rangle$ relation, then 
\begin{equation}
 {V - \langle V\rangle \over \sigma_V}
  = {L - \langle L\rangle\over\sigma_L}\ r_{VL} + n\, \sqrt{1 - r_{VL}^2},
 \label{selectV}
\end{equation}
and inserting this in the expression above yields 
\begin{equation}
 {\langle M_\bullet|L\rangle\over\sigma_\bullet} 
 = {\langle M_\bullet\rangle\over\sigma_\bullet} 
   + n\,{r_{\bullet V} - r_{\bullet L}r_{VL}\over \sqrt{1 - r_{VL}^2}}
   + {L - \langle L\rangle\over\sigma_L}\,r_{\bullet L}.
\end{equation}
The slope of this relation is still $a_{\bullet|L}$; because 
the selection was done at fixed $L$, the slope of the observed 
$\langle M_\bullet|L\rangle$ relation is not biased from its 
true value.  However, the zero-point may be strongly affected.  
The magnitude of the offset depends on how much of the 
$M_\bullet-\sigma$ relation is due to the $\sigma-L$ and 
$M_\bullet-L$ relations.  
If $n$ is positive (as Figure~\ref{sigmaLv} suggests) and 
$r_{\bullet V}>r_{\bullet L}r_{VL}$, then the observed relation 
predicts larger black hole masses for a given luminosity than 
would the true relation.  The associated 
$\langle M_\bullet|\sigma\rangle$ relation is obtained by 
transforming $L$ to $V$ using equation~(\ref{selectV}):
\begin{eqnarray}
 {\langle M_\bullet|V\rangle\over\sigma_\bullet} 
 &=& {\langle M_\bullet\rangle\over\sigma_\bullet} + 
   {V-\langle V\rangle\over\sigma_V}\, r_{\bullet V}\, 
    \left({r_{\bullet L}\over r_{\bullet V}r_{VL}} \right)\nonumber\\
 &&  + n\, r_{\bullet V}\,{1 - (r_{\bullet L}/r_{\bullet V}r_{VL}) 
                           \over \sqrt{1 - r_{VL}^2}}.
\end{eqnarray}
Both the slope and zero-point of this relation are affected, and 
the magnitude of the effect depends on whether or not 
$r_{\bullet L}>r_{\bullet V}r_{VL}$.  If the correlation between 
$M_\bullet$ and $L$ is stronger(weaker) than can be accounted for 
by the $M_\bullet-V$ and $V-L$ relations, then the observed 
$\langle M_\bullet|V\rangle$ has a steeper (shallower) slope than 
the true relation, and a shifted zero-point, such that it predicts 
smaller (larger) masses for a given $\sigma$ than the true relation.  
Note that this means it is possible for {\em both} the $L-$ and 
$\sigma-$based approaches to overestimate the true abundance of 
massive black holes.  

If the selection chooses objects with small $L$ for their $\sigma$, 
then the resulting relations are given by exchanging $L$s for $V$s 
in the previous two expressions.  In this case, 
$\langle V|L\rangle$ and $\langle M_\bullet|L\rangle$ both have 
biased slopes and zero points, whereas 
$\langle M_\bullet|\sigma\rangle$ has the correct slope but a 
shifted zero-point.  Once again, the sign and magnitude of the bias 
depends on how much of the $M_\bullet-$observable correlation is due 
to the other two correlations.  
For the choice of parameters used to make the Figures in the 
previous subsection,
 $r_{\bullet V}r_{VL} = 0.69 \approx r_{\bullet L}$:
the $M_\bullet-L$ relation is almost entirely a 
consequence of the other two correlations.  Thus, this toy 
model shows why the $\langle M_\bullet|\sigma\rangle$ relation 
shown in Figure~\ref{mockRels} was unbiased by the selection 
(which was done at fixed $\sigma$).

\section{Discussion}\label{sowhat}
Compared to the ENEAR and SDSS-B06 early-type galaxy samples, the 
$\sigma-L$ correlation in black hole samples currently available 
is biased towards large $\sigma$ for a given $L$ 
(Figures~\ref{sigmaLv}--\ref{sigmaLK}).  
If this is a selection effect rather than a physical one, then current 
determinations of the $M_\bullet-L$ and $M_\bullet-\sigma$ relations 
are biased.  We provided a simple toy model of the effect which shows 
how these biases depend on the intrinsic correlations and on the 
selection effect.  

To illustrate its use, we constructed simple models of the intrinsic 
correlation and of the selection effect.  We then identified a 
particular set of parameters which we showed can reproduce 
all the observed trends:  the correlations between $M_\bullet$ and 
$L$ or $\sigma$, the $\sigma-L$ correlation in black hole samples, 
as well as correlations between residuals from these fits 
(Figures~\ref{mockRels} and~\ref{SIMresids}) are all reproduced.  
We do not claim to have found the unique solution to this problem.  
However, if we have correctly modeled the selection effect, then our 
analysis suggests that the bias in the $\langle M_\bullet|\sigma\rangle$ 
relation is likely to be small, whereas the 
$\langle M_\bullet|L\rangle$ relation is biased towards having a 
steeper slope (Figure~\ref{mockRels}).  
This has the important consequence that estimates of black hole 
abundances which combine the $\langle M_\bullet|L\rangle$ relation 
and its scatter with an external determination of the distribution 
of $L$ will overpredict the true abundances of massive black holes.  
The analogous approach based on $\sigma$ is much less biased 
(Figure~\ref{mockCounts}).  

The choice of parameters which reproduces the selection-biased 
fits reported in Appendix~A has intrinsic distribution 
$\phi(M_\bullet)$ that is well described 
\begin{equation}
 \phi(M_\bullet)\,{\rm d}M_\bullet 
   = {\phi_*\over \Gamma(\alpha/\beta)}\, 
      \mu^\alpha\,\exp(-\mu^\beta)\,\beta\,{{\rm d}M_\bullet\over M_\bullet}
 \label{phiMbh}
\end{equation}
where $\phi_* = 0.002/{\rm Mpc}^{3}$, 
 $(\alpha,\beta) = (2.0,0.3)$ and $\mu = M_\bullet/M_\bullet^*$, 
with $M_\bullet^*$ a characteristic mass.  
This mass is related to the mean mass $10^{8}M_\odot$ by 
$\langle M_\bullet\rangle 
  = M_\bullet^*\,\Gamma[(\alpha+1)/\beta]/\Gamma(\alpha/\beta)$, 
so $M_\bullet^* = 10^{5}\,M_\odot$.  Since this $\phi(M_\bullet)$ 
is based on the SDSS-B06 luminosity and velocity functions, it is 
missing small bulges, so it is incomplete at small $M_\bullet$.  

The intrinsic correlations in the sample are 
\begin{equation}
 \Bigl\langle M_\bullet|M_r\Bigr\rangle = 9.16 - 0.41\,(M_r+22)\\
\end{equation}
and 
\begin{equation}
 \Bigl\langle M_\bullet|\log\sigma\Bigr\rangle 
  = 8.85 + 3.82\, \log\, (\sigma/200~{\rm km~s}^{-1})
\end{equation}
with scatter of 0.38 and 0.23~dex, respectively.  
A bisector fit to the $M_\bullet-L$ relation would have slope 
$(\sigma_\bullet/\sigma_L) (r_{\bullet L} + 1/r_{\bullet L})/2 = -0.62$; 
a similar fit to the $M_\bullet-V$ relation would have slope 
$(\sigma_\bullet/\sigma_V) (r_{\bullet V} + 1/r_{\bullet V})/2 = 4.49$.
Notice that the slopes satisfy 
$a_{\bullet|L}\approx a_{\bullet|V}\,a_{V|L}$.  
This is because, for this choice of parameters, the correlation 
between $M_\bullet$ and $L$ is almost entirely a consequence of 
the other two correlations.  

If we have correctly modelled the selection effect, and we do not 
claim to have done so, then our analysis suggests that the 
intrinsic joint distribution of $M_\bullet$, luminosity and 
$\sigma$ is similar to that between color, luminosity and $\sigma$:  
of the three pairwise correlations, the two correlations involving 
$\sigma$ are fundamental, whereas the third is not 
(Bernardi et al. 2005).  Since residuals from 
the $\langle\sigma|L\rangle$ relation are correlated with age 
(Bernardi et al. 2005), the correlations shown in the top right 
panels of Figures~\ref{SIMresids} and~\ref{BHresids} suggest that, 
at fixed luminosity, older galaxies tend to host more massive black 
holes than less massive galaxies.  Understanding why this is true may 
provide important insight into black hole formation and evolution.

\acknowledgements

This work is partially supported by NASA grant LTSA-NNG06GC19G, 
and by grants 10199 and 10488 from the Space Telescope Science 
Institute, which is operated by AURA, Inc., 
under NASA contract NAS 5-26555.

Funding for the SDSS and SDSS-II
has been provided by the Alfred P. Sloan Foundation, 
the Participating Institutions, the NSF, the US DOE, NASA, 
the Japanese Monbukagakusho, the Max Planck Society 
and the Higher Education Funding Council for England.  
The SDSS website is http://www.sdss.org/.

The SDSS is managed by the Astrophysical Research Consortium (ARC) 
for the Participating Institutions:  The American Museum of Natural 
History, Astrophysical Institute Postdam, the University of Basel, 
Cambridge University, Case Western Reserve University, 
the University of Chicago, Drexel University, Fermilab, 
the Institute for Advanced Study, the Japan Participation 
Group, the Johns Hopkins University, the Joint Institute for 
Nuclear Astrophysics, the Kavli Institute for Particle Astrophysics 
and Cosmology, the Korean Scientist Group, the Chinese Academy 
of Sciences (LAMOST), Los Alamos National Laboratory, 
the Max Planck Institute for Astronomy (MPI-A), 
the Max Planck Institute for Astrophysics (MPA), 
New Mexico State University, the Ohio State University, 
the University of Pittsburgh, the University of Portsmouth, 
Princeton University, the U.S. Naval Observatory, and the 
University of Washington.

\appendix
\section{The $\sigma-L$ correlation}
The main text compares the $\sigma-L$ relation for the bulk of 
the early-type galaxy population with that for the black hole 
sample.  This was done by using luminosities transformed to 
the SDSS $r-$band and scaled to $H_0 = 70$~km~s$^{-1}$~Mpc$^{-1}$.  
Specifically, we convert from B, V, R, I and K-band luminosities to 
SDSS $r-$band using 
\begin{displaymath}
 B-r = 1.25,\quad V-r=0.34 \quad r-R=0.27,\quad  r-I = 1.07,\quad {\rm and} \quad r-K = 2.8;
\end{displaymath}
these come from Fukugita et al. (1995) for S0s and Es.  
We describe the samples and required transformations below.

\begin{table}
\caption[]{Parameters of the black hole sample used in this work.  
           HR=H\"aring \& Rix (2004); MH = Marconi \& Hunt (2003).\\
           $M_r^{\rm HR}$ is from observed absolute magnitude; 
           $M_r^{\rm HR-M_{bulge}}$ is inferred from 
           $M_r^{\rm HR} -M_{\rm bulge}$ relation.   }
\small
\centering
\begin{tabular}{ccccccc}
 \hline &&&\\
  name & $\sigma^{\rm HR}$ & Log$_{10}$ M$_{\bullet}^{\rm HR}$ & $M_r^{\rm HR}$ & $M_r^{\rm HR-M_{bulge}}$ & $\sigma^{\rm MH}$ & $M_K^{\rm MH}$ \\
  & [km s$^{-1}]$ & [M$_{Sun}$] & [mag] & [mag] & [km s$^{-1}]$ & [mag] \\
\hline &&&\\

     M87 & 375 &   9.54 & $-$23.08 & $-$22.86 & 375 & $-$25.60 \\
 NGC3379 & 206 &   8.06 & $-$20.85 & $-$20.94 & 206 & $-$24.20 \\
 NGC4374 & 296 &   8.69 & $-$22.22 & $-$22.41 &$--$ & $--$     \\
 NGC4261 & 315 &   8.77 & $-$21.90 & $-$22.41 & 315 & $-$25.60 \\
 NGC6251 & 290 &   8.78 & $-$22.69 & $-$22.80 & 290 & $-$26.60 \\
 NGC7052 & 266 &   8.58 & $-$22.57 & $-$22.22 & 266 & $-$26.10 \\
 NGC4742 &  90 &   7.20 & $-$19.75 & $-$18.83 &  90 & $-$23.00 \\
 NGC0821 & 209 &   7.63 & $-$21.43 & $-$21.51 & 209 & $-$24.80 \\
  IC1459 & 323 &   9.46 & $-$22.37 & $-$22.22 & 340 & $-$25.90 \\
     M32 &  75 &   6.46 & $-$16.99 & $-$17.02 &  75 & $-$19.80 \\
 NGC2778 & 175 &   7.20 & $-$20.74 & $-$21.04 & 175 & $-$23.00 \\
 NGC3115 & 230 &   9.06 & $-$21.12 & $-$21.44 & 230 & $-$24.40 \\
 NGC3245 & 205 &   8.38 & $-$20.85 & $-$20.94 & 205 & $-$23.30 \\
 NGC3377 & 145 &   8.06 & $-$20.06 & $-$19.67 & 145 & $-$23.60 \\
 NGC3384 & 143 &   7.26 & $-$20.17 & $-$19.86 & 143 & $-$22.60 \\
 NGC3608 & 182 &   8.34 & $-$21.24 & $-$21.26 & 182 & $-$24.10 \\
 NGC4291 & 242 &   8.55 & $-$21.24 & $-$21.51 & 242 & $-$23.90 \\
 NGC4473 & 190 &   8.10 & $-$21.18 & $-$21.21 & 190 & $-$23.80 \\
 NGC4564 & 162 &   7.81 & $-$20.31 & $-$20.56 & 162 & $-$23.40 \\
 NGC4649 & 375 &   9.36 & $-$22.50 & $-$22.69 & 385 & $-$25.80 \\
 NGC4697 & 177 &   8.29 & $-$21.44 & $-$21.37 & 177 & $-$24.60 \\
 NGC5845 & 234 &   8.44 & $-$20.11 & $-$20.41 & 234 & $-$23.00 \\
 NGC7332 & 122 &   7.17 & $-$20.28 & $-$19.61 &$--$ & $--$     \\
 NGC7457 &  67 &   6.60 & $-$18.85 & $-$18.94 &  67 & $-$21.80 \\

\hline &&&\\
\end{tabular}
\label{tab:doubles} 
\end{table}
\normalsize

\begin{figure*}
\centering
\includegraphics[width=0.9\columnwidth]{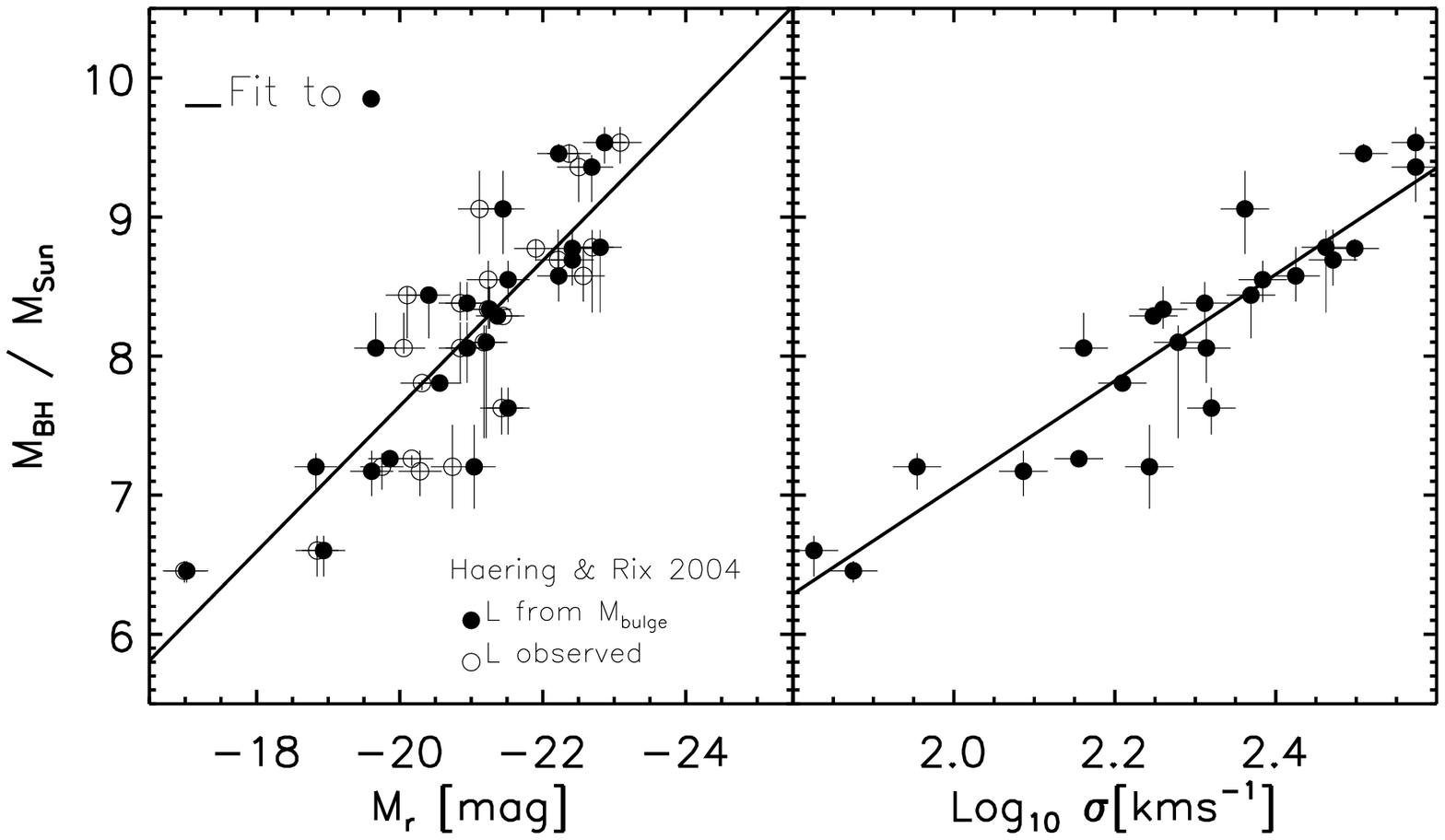}
 \caption{The $M_\bullet-L$ and $M_\bullet-\sigma$ relations 
          in the subset of the H\"aring \& Rix (2004) sample 
          which we describe in the main text.  Solid lines 
          show the fits to the filled circles given in 
          equations~(\ref{mbhLum}) and~(\ref{mbhSigma}).}
 \label{HRcorr}
\end{figure*}

\begin{figure*}
\centering
\includegraphics[width=0.9\columnwidth]{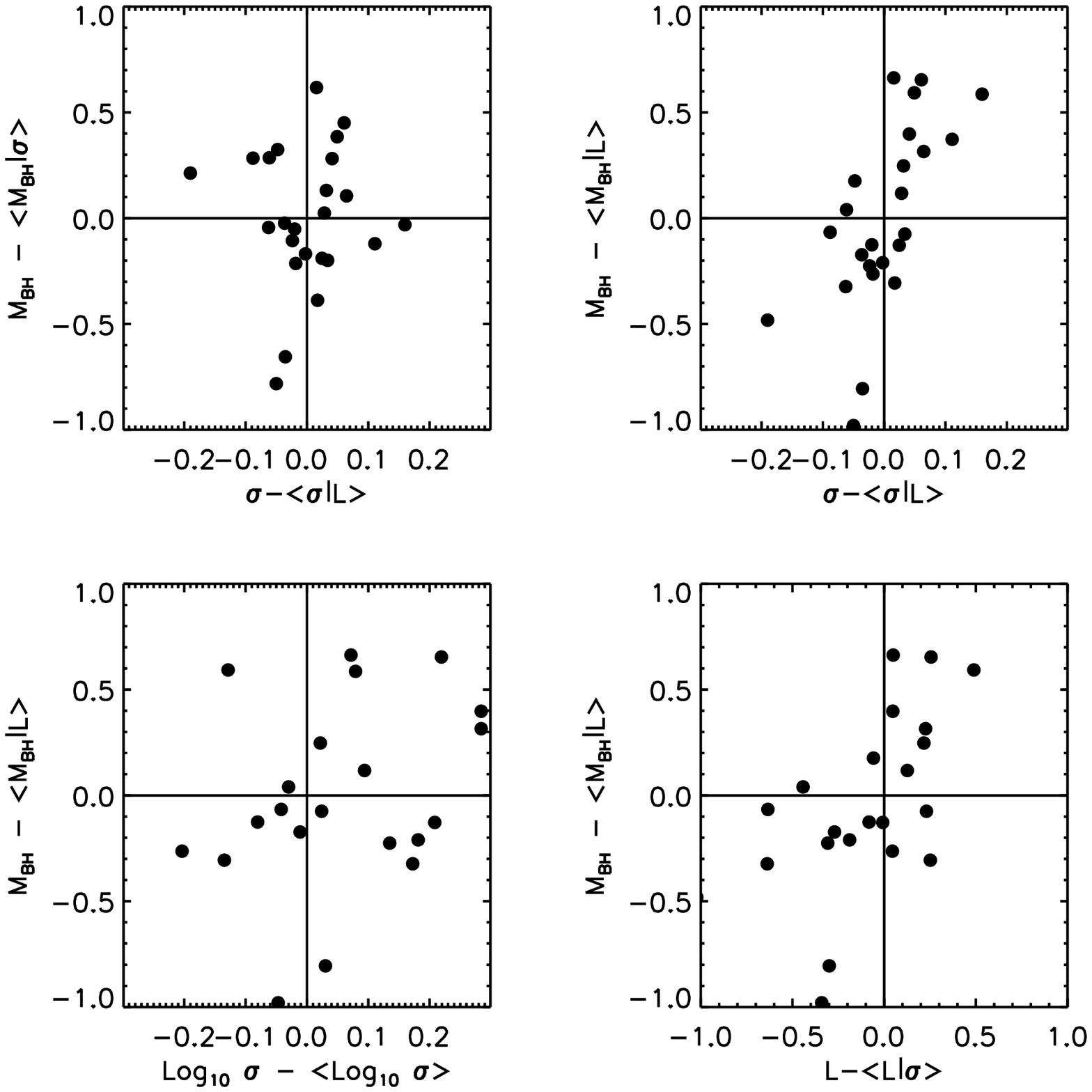}
 \caption{Residuals from the $M_\bullet-\sigma$ relation do not 
          correlate with residuals from the $\sigma-L$ correlation 
          (top left); whereas residuals from the $M_\bullet-L$ 
          relation do (top right).  Residuals from the 
          $M_\bullet-L$ relation also correlate with residuals 
          from the $L-\sigma$ relation (bottom right), but they 
          correlate weakly if at all with $\sigma$ (bottom left). }
 \label{BHresids}
\end{figure*}

\subsection{The black hole sample}
The main properties of the black hole sample shown in Figure~1 
are listed in Table~1.  The sample was compiled as follows.  
We selected all objects classified as E or S0 
by H\"aring \& Rix (2004) in their compilation of black holes.  
We excluded NGC~4342 as they recommend, as well as NGC~1023 which 
Ferrarese \& Ford (2005) say should be excluded because it is not 
sufficiently well resolved.  
H\"aring \& Rix provide estimates of the bulge luminosities and 
masses of these objects; these assume $H_0=70$~km~s$^{-1}$~Mpc$^{-1}$.  
For consistency we also rescaled the black hole masses from 
$H_0=80$~km~s$^{-1}$~Mpc$^{-1}$ to $H_0=70$~km~s$^{-1}$~Mpc$^{-1}$.
We convert from the V, R and I-band luminosities the report to 
SDSS $r$ using the transformations given above.  
We use the bulge masses to provide an alternative estimate of the 
bulge luminosities by setting 
\begin{equation}
 M_r = -22.34 - {\log_{10}(M_{\rm bulge}/M_\odot) - 11.52\over 0.492};
\end{equation}
this was obtained from a linear regression of $M_r$ on 
$M_{\rm bulge}$.  
Table~1 also lists estimates of the K-band near-infrared luminosity 
from Marconi \& Hunt (2003) which are shown in Figure~2.

A word on the velocity dispersions is necessary.
H\"aring \& Rix (2004) and Marconi \& Hunt (2003) use the velocity 
dispersions provided by Kormendy \& Gebhardt (2001) (two galaxies in
Marconi \& Hunt have a different value). The two estimates correspond to 
different apertures ($r_e$ and central, which probably means $r_e/8$).  
Tremaine et al. (2002) provide a detailed discussion of aperture 
effects.  However, it is not clear that their assumptions about 
the scale dependence of the stellar velocity dispersion are 
consistent with more recent measurements (Cappellari et al. 2006).  
Since the difference in aperture is not sufficient to account 
for the discrepancy between black hole samples and the normal 
early-type galaxy samples we have not made any correction for this 
difference.  

The $M_\bullet-L$, $M_\bullet-\sigma$, and $\sigma-L$ 
correlations in the H\"aring \& Rix (2004) sample, with 
luminosities estimated from $M_{\rm bulge}$ are 
\begin{equation}
 \Bigl\langle \log M_{\bullet}|M_r\Bigr\rangle 
  = (8.57\pm 0.10) - {(1.30\pm 0.10)\over 2.5}\,(M_{r}+22)
 \label{mbhLum}
\end{equation}
with scatter of 0.33~dex, 
\begin{equation}
 \Bigl\langle \log M_{\bullet}|\log \sigma\Bigr\rangle 
  = (8.21 \pm 0.05) + 
 (3.83 \pm 0.10) \,\log\left({\sigma\over 200\,{\rm kms}^{-1}}\right)
 \label{mbhSigma}
\end{equation}
with intrinsic scatter of 0.22~dex, and 
\begin{equation}
 \Bigl\langle \log \sigma|M_r\Bigr\rangle 
        = (2.41\pm 0.10) - {(0.34\pm 0.03)\over 2.5}\,(M_{r} + 22)
 \label{bhLS}
\end{equation}
(Tundo et al. 2006).  
Figure~\ref{HRcorr} shows the $M_\bullet-L$ and $M_\bullet-\sigma$ 
relations and these fits.  The dotted line in Figure~\ref{sigmaLv} 
of the main text shows equation~(\ref{bhLS}).  

Figure~\ref{BHresids} shows that residuals from the 
$M_\bullet-\sigma$ relation do not correlate with residuals from 
the $\sigma-L$ correlation (top left), whereas residuals from the 
$M_\bullet-L$ relation do (top right).  
Residuals from the $M_\bullet-\sigma$ relation 
do not correlate with $\sigma$, $L$ or $L-\sigma$ residuals, so we 
have not shown these correlations here.  However, residuals from 
the $M_\bullet-L$ relation do correlate with residuals from the 
$L-\sigma$ relation (bottom right) and they may also correlate 
weakly with $\sigma$ itself (bottom left).  

The top right panel suggests that, at fixed luminosity, objects 
with larger $M_\bullet$ tend to have larger $\sigma$.  
The bottom right panel suggests that these objects also tend to 
be faint for their $\sigma$.  
But whether or not these are entirely selection effects 
is unclear; 
this is because the relations~(\ref{mbhLum})--(\ref{bhLS}) are a 
combination of an intrinsic correlation and a selection bias, so 
that correlations between residuals are also combinations of true 
and selection-biased relations.

\subsection{Fits to early-type galaxy samples}
The velocity dispersions for the early-type galaxy population 
were scaled from their measured values to those they are expected 
to have within a circular aperture of radius $r_e/8$.  
For ENEAR, when the observed redshift is used as the distance 
indicator (i.e. no correction is made for peculiar velocities), 
\begin{equation}
 \Bigl\langle \log \sigma|M_r\Bigr\rangle 
        = (2.300\pm 0.110) - {(0.255\pm 0.018)\over 2.5}\,(M_{r} + 22).
 \label{enearz}
\end{equation}
Correcting for peculiar velocities leads to a biased relation, 
because the correction makes $L$ depend on $\sigma$.  
This biased relation is 
\begin{equation}
 \Bigl\langle \log \sigma|M_r\Bigr\rangle 
        = (2.316\pm 0.100) - {(0.318\pm 0.011)\over 2.5}\,(M_{r} + 22).
 \label{eneard}
\end{equation}
The galaxies in the SDSS-B06 are sufficiently distant that the 
SDSS-B06 $\sigma-L$ relation is not affected by peculiar velocities.  
It is 
\begin{equation}
 \Bigl\langle \log \sigma|M_r\Bigr\rangle 
        = (2.290\pm 0.080) - {(0.250\pm 0.007)\over 2.5}\,(M_{r} + 22).
 \label{sdss}
\end{equation}
Expressions~(\ref{enearz})---(\ref{sdss}) are shown in 
Figure~\ref{sigmaLv} of the main text.  They are taken from 
Bernardi (2006), where a discussion of the systematics in the 
magnitudes and velocity dispersion in the SDSS database can 
be found.  

\subsection{Analogous quantities in the K-band}
The main text also studied the $\sigma-L_{\rm K}$ relation in 
the black hole compilation of Marconi \& Hunt (2003).  
This relation is 
\begin{equation}
 \Bigl\langle \log \sigma|M_{\rm K}\Bigr\rangle 
        = (2.373\pm 0.110) - {(0.278\pm 0.022)\over 2.5}\,(M_{\rm K}+24.8),
 \label{marconiHunt}
\end{equation}
where the fit is made only to the objects which are also in 
the H\"aring \& Rix (2004) compilation.  
Matching SDSS-B06 to 2MASS, and so estimating $\sigma-L_{\rm K}$
for early-types yields 
\begin{equation}
 \Bigl\langle \log \sigma|M_{\rm K}\Bigr\rangle 
        = (2.304\pm 0.100) - {(0.250\pm 0.008)\over 2.5}\,(M_{\rm K}+24.8).
 \label{sdss2mass}
\end{equation}

\end{document}